\documentclass[aps,prl,times,twocolumn,amsmath,amssymb,superscriptaddress,floatfix,showpacs]{revtex4}

\usepackage{color}
\usepackage{graphicx}
\usepackage{bbold}
\usepackage{dsfont}

\begin{document}


\newcommand{\hexa}{\;
\pspicture(0,0.1)(0.35,0.6)
\psset{unit=0.75cm}
\pspolygon(0,0.15)(0,0.45)(0.2598,0.6)(0.5196,0.45)(0.5196,0.15)(0.2598,0)
\psset{linewidth=0.12,linestyle=solid}
\psline(0,0.15)(0,0.45)
\psline(0.2598,0.6)(0.5196,0.45)
\psline(0.5196,0.15)(0.2598,0)
\endpspicture\;}

\newcommand{\hexb}{\;
\pspicture(0,0.1)(0.35,0.6)
\psset{unit=0.75cm}
\psset {linewidth=0.03,linestyle=solid}
\pspolygon[](0,0.15)(0,0.45)(0.2598,0.6)(0.5196,0.45)(0.5196,0.15)(0.2598,0)
\psset{linewidth=0.12,linestyle=solid}
\psline(0.2598,0.6)(0,0.45)
\psline(0.5196,0.12)(0.5196,0.45)
\psline(0.2598,0)(0,0.15)
\endpspicture\;}

\newcommand{\smallhex}{ \;
\pspicture(0,0.1)(0.2,0.3)
\psset{linewidth=0.03,linestyle=solid}
\pspolygon[](0,0.0775)(0,0.225)(0.124,0.3)(0.255,0.225)(0.255,0.0775)(0.124,0)
\endpspicture
\;}


\title{Quantum Ice : a quantum Monte Carlo study}


\author{Nic Shannon}
\affiliation{H. H. Wills Physics Laboratory, University of Bristol, Tyndall Avenue, Bristol BS8 1TL, UK.}

\author{Olga Sikora}
\affiliation{H. H. Wills Physics Laboratory, University of Bristol, Tyndall Avenue, Bristol BS8 1TL, UK.}

\author{Frank Pollmann}
\affiliation{Max-Planck-Institut f{\"u}r Physik komplexer Systeme, 01187 Dresden, Germany}

\author{Karlo Penc}
\affiliation{Research Institute for Solid State Physics and Optics, H-1525 Budapest, P.O.B. 49, Hungary.}

\author{Peter Fulde}
\affiliation{Max-Planck-Institut f{\"u}r Physik komplexer Systeme, 01187 Dresden, Germany}
\affiliation{Asia Pacific Center for Theoretical Physics, Pohang, Korea}


\date{\today}


\begin{abstract}
Ice states, in which frustrated interactions lead to a macroscopic ground-state 
degeneracy, occur in water ice, in problems of frustrated charge order 
on the pyrochlore lattice, and in the family of rare-earth magnets collectively 
known as spin ice.
Of particular interest at the moment are ``quantum spin ice'' materials, 
where large quantum fluctuations may permit tunnelling 
between a macroscopic number of different classical ground states.
Here we use zero-temperature quantum Monte Carlo simulations
to show how such tunnelling can lift the degeneracy of a spin 
or charge ice, stabilising a unique ``quantum ice'' ground state 
--- a quantum liquid with excitations described by the 
Maxwell action of 3$+$1-dimensional quantum electrodynamics.  
We further identify a competing ordered ``squiggle'' state, and show how 
both squiggle and quantum ice states might be distinguished in neutron 
scattering experiments on a spin ice material.
\end{abstract}




\maketitle


Ice is one of the strangest substances known to man.  
In the common forms of water ice, protons occupy the space between tetrahedrally 
coordinated oxygen ions, and each oxygen obeys the ``ice rule'' constraint of forming 
two long and two short bonds with neighbouring protons~\cite{bernal33,pauling35}.     
It was quickly realised that these ice rules did not select a single, unique
proton configuration~\cite{bernal33}, but rather a vast manifold of  
classical ground states, with an entropy {\it per water molecule}
of $s_0 \approx {\text k}_{\sf B} \log (3/2)$~\cite{pauling35}.
This prediction proved to be in good agreement with measurements 
of entropy at low temperatures~\cite{giauque36}, but stands in clear violation of the 
third law of thermodynamics --- at zero temperature we expect 
water ice to be described by a single, unique, ground state wave function. 


The same ``ice'' rules, and the same extensive 
ground state degeneracy, arise in  
(i) problems of frustrated charge~\cite{anderson58,fulde02} and orbital~\cite{chern01} order; 
(ii) proton-bonded ferroelectrics~\cite{youngblood80};
(iii) statistical descriptions of polymer melts~\cite{kondev98};
and 
(iv) a family of rare-earth magnets collectively known as 
``spin-ice''~\cite{harris97,ramirez99,bramwell01,gardner10}.   
In each case, the ice rules have non-trivial consequences for the properties of the 
system, notably an algebraic decay of correlations~\cite{youngblood80, huse03, henley05, henley10} 
and excitations with ``fractional'' character~\cite{bernal33, kondev98, fulde02, ryzhkin05, castelnovo08}.
These exotic features of the ice state have been extremely well characterized 
in spin ice, where the algebraic decay of correlation functions is visible 
as ``pinch points'' in the magnetic structure factor~\cite{fennell09}, 
and the fractional excitations have the character of magnetic 
monopoles~\cite{bramwell08, morris09, kadowaki09, jaubert09}. 


\begin{figure}[ht]
\centering
\includegraphics[width=0.9\columnwidth]{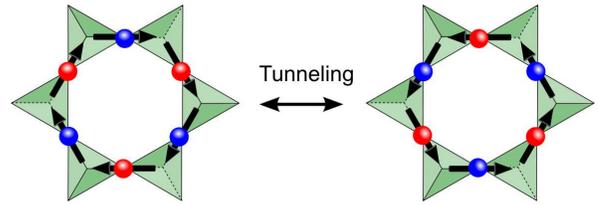}
\caption{
\footnotesize{
(Color online) 
Quantum mechanics enters ice physics through the tunnelling of the system from one
ice configuration to another.
The leading tunnelling matrix element for spin ice is the reversal 
of a set of Ising spins with closed circulation on 6-link hexagonal plaquette.
}}
\label{fig:tunneling_matrix_element}
\end{figure}
  

All of these systems beg the obvious question --- how is the degeneracy of the ice
manifold lifted at zero temperature~?  
The simplest way for an ice to recover a unique ground state at zero temperature
is for it to order.
This is exactly what happens in KOH-doped water ice, where 
the protons order below 70K~\cite{kawada72}.
However in many spin-ice materials, {\it no} order is observed~\cite{gardner10}.  
This raises the intriguing possibility that there might exist a zero temperature 
 ``quantum ice'' state, in which a single quantum mechanical ground state is formed 
through the coherent superposition of an exponentially large number of classical 
ice configurations.   
Such a state could have a vanishing entropy at zero temperature, and so satisfy the 
third law of thermodynamics, without sacrificing the algebraic correlations and fractional 
excitations (magnetic monopoles) associated with the degeneracy of the ice states.


In this Letter we use zero-temperature quantum Monte Carlo simulations to establish the 
ground state of the minimal microscopic model of a charge or spin ice with tunnelling between 
different ice configurations.    
We find that the ground state is a quantum liquid, with an emergent $U(1)$ gauge symmetry, 
and excitations described by the Maxwell action of 3$+$1-dimensional quantum electrodynamics.
This state is the exact, quantum, analogue of the spin-liquid phase realised
in ``classical'' spin ices such as Dy$_2$Ti$_2$O$_7$, and exhibits the same
magnetic monopole excitations.
We also explore how quantum effects in this novel liquid modify the ``pinch--point'' 
singularities seen in neutron scattering experiments on spin ices.


\begin{figure}[ht!!!]
\centering
\includegraphics[width=0.8\columnwidth]{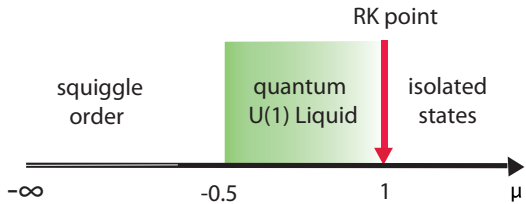}
\caption{
\footnotesize{
(Color online) 
Ground state phase diagram of the quantum ice model [Eq.~(\ref{eq:H})] 
as a function of the ratio $\mu$ of kinetic to potential energy.
The 3D ``quantum ice'' point $\mu=0$ exists deep within an extended quantum liquid phase 
with deconfined fractional excitations and algebraic decay of correlation functions.  
}}
\label{fig:phase_diagram}
\end{figure}


The best systems in which to look for a quantum ice are those which are able to tunnel 
from one ice configuration to another.
In water ice, in the absence of mobile ionic defects~\cite{bjerrum52}, 
this tunnelling occurs through the collective hopping of protons on a 6-link loop.
In spin ice it is the cyclic exchange of Ising spins on a hexagonal plaquette~\cite{bramwell98}, 
illustrated in Fig.~\ref{fig:tunneling_matrix_element}.
In both cases the ice rules can be written as a compact lattice 
U(1)-gauge theory in which the displacement of protons --- or orientation of 
magnetic moments --- are associated with a fictitious magnetic field 
$\mathbf{B} = \nabla \times \mathbf{A}$, 
in the Coulomb gauge $\nabla\cdot{\bf A}=0$~\cite{huse03,henley05}.   
Tunnelling between different ice configurations introduces dynamics in the 
gauge field $\mathbf{A}$, and the minimal description of a quantum ice is
the Maxwell action of conventional quantum electromagnetism
\begin{eqnarray}
\mathcal{S}=\int d^3xdt\left[ \mathbf{E}^2 - c^2 \mathbf{B}^2 \right],
\label{eq:maxwellS}
\end{eqnarray}
where $c$ is the effective speed of light and, in the absence of electric charges,
$\mathbf{E} = -\partial \mathbf{A}/\partial t$.  
It follows directly from Eq.~(\ref{eq:maxwellS}) 
that correlation functions have dipolar character, and local defects in an ice 
configuration act like deconfined magnetic monopoles~\cite{moessner03,hermele04,castro-neto06}.


\begin{figure}[ht!!!]
\centering
\includegraphics[width=0.65\columnwidth]{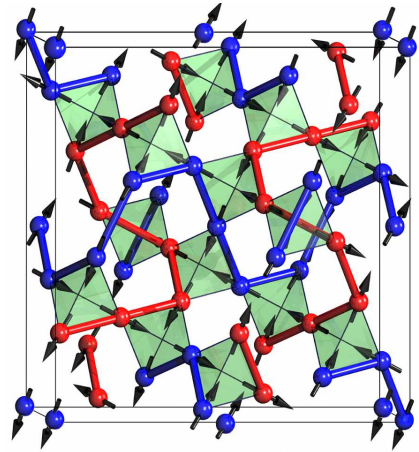}\caption{
\footnotesize{
(Color online) 
The ice configuration possesing the most flippable plaquettes 
is the ``squiggle'' state, shown here within a 40-site tetragonal cell.
Arrows show the displacement of protons within $Ic$ water ice or, 
equivalently, the orientation of spins in spin ice.    
The squiggle state posses a net ``magnetic'' flux, 
\mbox{$\vec\phi = \vec\phi_{\sf squiggle}$}, 
orientated along a [100] axis of the crystal.
In a spin ice material this would correspond to a net magnetisation 
of $1/5$ the value at saturation, directed along a [100] axis. 
}}
\label{fig:squiggle}
\end{figure}


Field-theoretical arguments alone cannot resolve 
whether the quantum $U(1)$ liquid described by Eq.~(\ref{eq:maxwellS}) 
is realised in an ``ice'' material, or realistic microscopic model.
Encouragingly, evidence supporting the existence of such a phase has been 
found in quantum Monte Carlo simulations of strongly-interacting hard-core 
Bosons on a pyrochlore lattice~\cite{banerjee08}.
However, these simulations are restricted to temperatures comparable 
with the degeneracy temperature of the ice manifold, and so are mute
as to the zero-temperature ground state.
In this article, we use zero-temperature Green's function Monte Carlo 
(GFMC) simulation techniques~\cite{GFMC} to provide concrete evidence for 
the existence of a quantum $U(1)$ liquid ground state in a microscopic lattice 
model of a quantum ice.


The model we consider was first introduced by 
Hermele~{\it et al.}~\cite{hermele04} as an effective Hamiltonian for
an easy-axis antiferromagnet on a pyrochlore lattice.
It is defined by the Hamiltonian
\begin{eqnarray}
\mathcal{H}_\mu &=&  
   -\sum_{\sf plaq.} \left[
   |\! \circlearrowright \rangle\langle \circlearrowleft\! | + 
   |\! \circlearrowleft \rangle\langle \circlearrowright\! | \right]
\nonumber\\&&
   + \mu  \sum_{\sf plaq.} \left[
   |\! \circlearrowright \rangle\langle \circlearrowright\! | + 
  |\! \circlearrowleft \rangle\langle \circlearrowleft\! | \right]
\label{eq:H}
\end{eqnarray}
acting on all possible (spin) ice configurations..
The first term in Eq.~(\ref{eq:H}) describes tunnelling from one
ice configuration to another, where $ |\! \circlearrowright \rangle$ should be 
understood as a closed circulation of $\mathbf{B}$ on a ``flippable'' hexagonal 
plaquette [cf. Fig.~\ref{fig:tunneling_matrix_element}].
The sum $\sum_{\sf plaq.}$ runs over all such plaquettes in the lattice.  
The additional potential energy term $\mu$ counts the number of flippable 
plaquettes in a given ice configuration, and renders the model
exactly soluble for $\mu=1$~\cite{rokhsar88}.
All energies are measured in units of the tunnelling matrix element
between ice configurations.
For $\mu=0$, Eq.~(\ref{eq:H}) is the minimal microscopic model for
a 3D quantum ice.  


\begin{figure}[ht!!!]
\centering
\includegraphics[width=0.95\columnwidth]{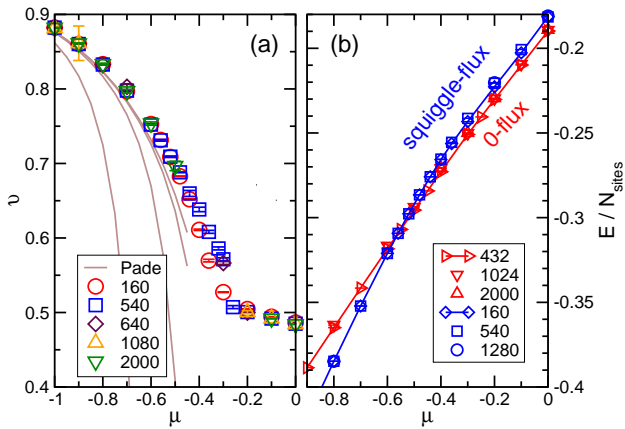}~~~
\caption{
\footnotesize{
(Color online) 
Evidence for a phase transition out of the squiggle state.
(a) Melting of squiggle order within the squiggle flux sector 
$\vec\phi = \vec\phi_{\sf squiggle}$.
The relative number of flippable plaquettes $\nu =N_{\sf f}(\mu)/N_{\sf f}(-\infty)$
is calculated using quantum Monte Carlo simulations 
for clusters of $160$ to $1080$ sites.  
The solid lines show different Pad\'e approximants 
to the series expansion about a perfectly-ordered squiggle state.  
(b) Ground state level crossing between the flux sector associated with the squiggle 
state, and the zero-flux sector associated with the quantum $U(1)$ liquid.  
The ground state energy per site is calculated for clusters of 432, 1024 and 2000 
sites (squiggle-flux sector) and 160, 540 and 1280 sites (zero-flux sector), using 
quantum Monte Carlo simulation. 
A clear crossing is observed for $\mu = -0.50\pm0.03$.  
}}
\label{fig:Nf_mu}
\label{fig:energy_crossing}
\end{figure}


We have previously used GFMC simulation to establish the existence of a quantum $U(1)$ 
liquid in the quantum dimer model on a diamond lattice~\cite{sikora09, sikora11}.  
The quantum ice model Eq.~\ref{eq:H} differs from this 
only in that the Hamiltonian acts on fully-packed loop, rather than dimer 
coverings of the lattice [c.f.~Ref.~\onlinecite{jaubert11}].   
We can therefore solve it using the methods set out in Ref.~\cite{sikora11}.
We consider clusters with periodic boundary conditions, and make extensive use of the 
fact that the ``magnetic'' flux  $\phi = \int d\mathbf{S} \cdot\mathbf{B}$ through {\it any} 
periodic boundary is a conserved quantity.   
This makes it possible to define a series of flux quantum numbers 
$\vec{\phi} = (\phi_x,\phi_y,\phi_z)$ for each cluster~\cite{sikora11}.  
Our findings are summarised in Fig.~\ref{fig:phase_diagram}.


For $\mu \to - \infty$, the ground state of Eq.~(\ref{eq:H}) is the ice configuration which 
maximizes the number of flippable plaquettes.
This is the 60-fold degenerate ``squiggle'' configuration shown in 
Fig.~(\ref{fig:squiggle}).
It is ordered, and therefore exhibits Bragg peaks in diffraction 
experiments~\cite{supplementary}.
A good measure of  squiggle order is the relative density of flippable plaquettes 
\mbox{$\nu =N_{\sf f}(\mu)/N_{\sf f}(-\infty)$}.  
In Fig.~\ref{fig:Nf_mu}(a), we compare GFMC calculations of $\nu$ for a series 
of finite-size clusters, carried out in the squiggle flux sector, with Pad\'e approximants 
to a series expansion in $1/\mu$ about perfect squiggle order.
The agreement between the two calculations is essentially perfect for $\mu < -0.75$, 
and the marked suppression in the number of flippable plaquettes for $\mu \gtrsim -0.3$ 
is strongly suggestive of the melting of squiggle order.   
However, perturbation theory about the exactly soluble ``RK'' point $\mu=1$ dictates that the ground 
state of Eq.~(\ref{eq:H}) should be in the zero-flux sector for $\mu \to 1$~\cite{hermele04,sikora11}.  
And in fact the gentle demise of squiggle order is pre-empted by a ground state level crossing between 
the squiggle and zero-flux sectors at $\mu = -0.50\pm0.03$, shown in Fig.~\ref{fig:energy_crossing}(b).  


\begin{figure}[ht!!!]
\centering
\includegraphics[width=0.95\columnwidth]{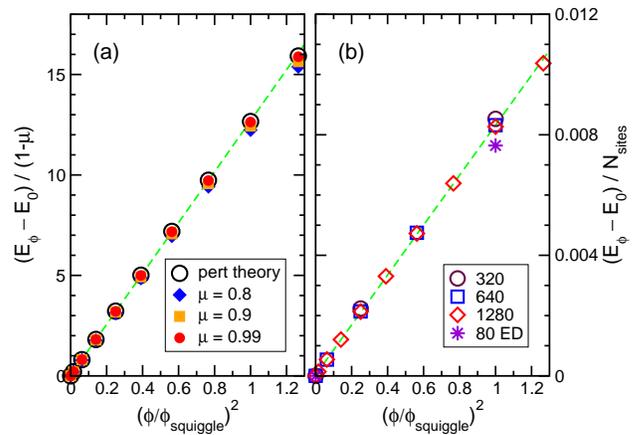}
\caption{
\footnotesize{
(Color online) 
Evidence for the existence of a quantum $U(1)$ liquid.  
(a) Flux dependence of finite-size energy gaps \mbox{$E_\phi-E_0$} for 
parameters bordering the RK point, $\mu=1$.  
Results are plotted for quantum Monte Carlo simulation of an 1280-site cluster (solid points) 
and perturbation theory about the RK point (open black circles).
The dashed line indicates the scaling expected for a quantum $U(1)$ liquid, 
following Eq.~(\ref{eq:maxwellS}). 
(b) Flux dependence of finite-size energy gaps 
at the \mbox{quantum-ice} point, $\mu=0$.  
Results are obtained using quantum Monte Carlo simulation for 320-site, 640-site and 
1280-site clusters, and in exact diagonalization for an 80-site cluster.
The dashed line indicates the scaling expected for a quantum $U(1)$ liquid. 
}}
\label{fig:RK_point}
\label{fig:ice_point}
\end{figure}


These results are consistent with a first order phase transition out of the squiggle state 
at \mbox{$\mu = -0.5$}, but do not yet confirm the existence of the quantum $U(1)$ 
liquid we are seeking.
Fortunately Eq.~(\ref{eq:maxwellS}) also makes specific predictions for how the quantum $U(1)$ 
liquid phase evolves out of the RK point at \mbox{$\mu=1$}~\cite{moessner03,hermele04,sikora11}.
Specifically, the finite-size energy gaps $E_\phi - E_0$ should grow as
$E_\phi-E_0  \sim c^2 \phi^2/L$
where $E_\phi$ is the energy of the ground state with flux $\vec{\phi}$
and $c^2 \propto 1 - \mu +\ldots$\cite{hermele04,sikora11}.
In Fig.~\ref{fig:RK_point}(a), we present simulation results for $E_\phi - E_0$ for $\mu\lesssim 1$.
We find good agreement between GFMC simulations and perturbation theory about the RK 
point~\cite{hermele04,sikora11}, and a near-perfect collapse of both data sets according 
to the prediction of Eq.~(\ref{eq:maxwellS}).
This confirms the existence of a quantum $U(1)$ liquid bordering the RK point, 
as proposed in~\cite{hermele04}.  


However so far as real materials are concerned, the most interesting point in the parameter 
space is the ``quantum ice'' point $\mu = 0$.
Does this also conform to the behaviour expected of a quantum $U(1)$-liquid ?
In Fig.~\ref{fig:ice_point}(b) we present simulation results for the finite-size 
energy gaps of Eq.~(\ref{eq:H}) at $\mu=0$.
We extract the leading dependence on system size by plotting 
\mbox{$(E_\phi-E_0)/N_{\sf sites}$} as a function of $(\phi/\phi_{\sf squiggle})^2$.  
Once again, the collapse of the data is excellent, confirming the existence 
of a quantum $U(1)$-liquid.  


The ground state phase diagram of the quantum ice model Eq.~(\ref{eq:H}) 
is summarised in Fig.~\ref{fig:phase_diagram}.
For \mbox{$\mu < -0.5$}, the ground state is the complex 
``squiggle'' order shown in Fig.~\ref{fig:squiggle}.  
For $\mu > -0.3$ we find unambiguous evidence for a quantum liquid phase
which is well-described by the $U(1)$ (lattice) gauge theory of 
quantum electromagnetism, and so will exhibit both algebraic decay 
of correlations and deconfined fractional excitations~\cite{moessner03,hermele04}.
This quantum $U(1)$ liquid phase terminates in an RK point at $\mu=1$, 
and therefore the ``quantum ice'' point $\mu = 0$ lies deep within it.
For $-0.5 < \mu < -0.3$ simulation results depend in detail on the geometry 
of the cluster chosen.  
However we tentatively conclude that a single, first order phase transition
takes place between the squiggle and $U(1)$-liquid phases for $\mu=-0.5$.
This phase diagram should be contrasted with those for bosonic~\cite{shannon04}
and fermionic~\cite{pollmann06} quantum ice models on the 2D square lattice, 
where all phases are ordered and confining for $\mu < 1$.
In short --- 2D quantum ice models are ordered and confining, but 
the 3D quantum ice model solved here is not.


So far as electronic charge ices are concerned, this work should be regarded as a 
``warm-up'' exercise, since Eq.~(\ref{eq:H}) does not allow for the spin or Fermi 
statistics of the electrons~\cite{pollmann06,sikora11,poilblanc08}. 
%
%
Similarly, the application of these ideas to hexagonal water ice depends crucially on 
the generalisation to a different lattice, and the role of more general ionic defects~\cite{bjerrum52}.
However the model we have solved may give a good account of ``quantum spin ice'' 
materials such as 
Tb$_2$Ti$_2$O$_7$~\cite{gardner99,gingras00,molavian07,molavian-arXiv}, 
Pr$_2$Sn$_2$O$_7$~\cite{zhou08,onoda10}
and 
Yb$_2$Ti$_2$O$_7$~\cite{ross11,thompson11,chang11}, 
where quantum fluctuations of magnetic moments provide a route to 
tunnelling between spin ice states.
%
%
And in this context it is interesting to ask how a quantum ice might be distinguished
in experiment on a spin ice material ?

\begin{figure}[ht!!!]
\centering
\includegraphics[width=\columnwidth]{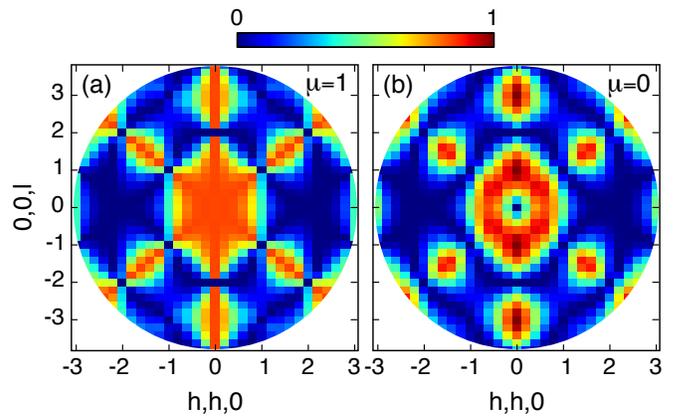}
\caption{
\footnotesize{
(Color online) 
(a)  Static structure factor $S({\bf q})$ in the spin-flip channel for a classical 
spin ice, as measured by Fennel {\it et al.}~\cite{fennell09}, calculated here 
from the microscopic model Eq.~(\ref{eq:H}), with $\mu=1$.  
Results are plotted in the $(h,h,l)$ plane, and show the ``pinch-point'' 
structure associated with $1/r^3$ dipolar correlations in $3$ dimensions.
(b) Equivalent static structure factor $S({\bf q})$ 
for a quantum spin ice described by Eq.~(\ref{eq:H}) with $\mu=0$. 
In this case, correlations between spins show the $1/r^4$ behaviour 
characteristic of dipoles in $3+1$ dimensions, and the pinch-points are 
eliminated.
All simulations were performed for a cubic cluster with 2000 lattice sites.
}}
\label{fig:Sq}
\end{figure}

The signal feature of a classical ice state is the presence of ``pinch point'' 
singularities in the static structure factor $S(q)$~\cite{fennell09}.  
These reflect the fact that the spins or charges which make up the ice state
have correlations of 3-dimensional dipolar form~\cite{youngblood80,huse03,henley05}.
In contrast, in a quantum ice, static correlations take on the form of dipoles in $3+1$ 
dimensions~\cite{hermele04,henley10} and as a result, the pinch-point singularities in $S(q)$ 
are eliminated.
To illustrate this, in Fig.~\ref{fig:Sq} we present GFMC simulation results 
for $S(q)$ for a quantum spin ice, calculated directly from Eq.~(\ref{eq:H}).  
Results for the RK point $\mu=1$, where the correlations are classical,  
clearly show pinch points at reciprocal lattice vectors.
However at the quantum ice point, $\mu=0$, there is a marked suppression 
of spectral weight around the same reciprocal lattice vectors.
As a consequence, the angle-integrated structure behaves as 
\mbox{$S(|{\bf q}| \to 0) \propto |{\bf q}|$}, while in a classical spin ice 
\mbox{$S(|{\bf q}| \to 0) \to const.$}    
This linear-$|{\bf q}|$ behaviour at small $|{\bf q}|$ is consistent with published 
results for Pr$_2$Sn$_2$O$_7$~\cite{zhou08}.
It might also be interesting to reexamine elastic neutron scattering data on 
other pyrochlore antiferromagnets in the light of these results~\cite{lee02}.  
These issues will be explored further elsewhere~\cite{benton11}.


The authors are pleased to acknowledge helpful conversations with Owen Benton, 
Michel Gingras, Chris Henley and Roderich Moessner.
This work was supported by EPSRC Grants EP/C539974/1 and EP/G031460/1, 
and Hungarian OTKA Grant K73455.  
KP, OS and NS gratefully acknowledge the hospitality of the guest program of 
MPI-PKS Dresden, where much of this work was carried out.



\end{document}